\documentclass[jkps,preprint,fleqn,showpacs,showkeys]{revtex4}
\usepackage{graphicx}
\usepackage{amssymb}
\usepackage{amsmath}
\usepackage{bm}
\begin{document}
\newcommand{\bea}{\begin{eqnarray}}
\newcommand{\eea}{\end{eqnarray}}
\newcommand{\wt}{\widetilde}
\setcounter{page}{0}
\title[]{Gravitational wave from cosmic inflation in a gravity with two small four-derivative corrections}
\author{Chae-min \surname{Yun}}
\affiliation{Department of Physics, Kyungpook National University, Taegu, Korea}
\email{clair.yun@gmail.com}

\begin{abstract}
 We investigate a model of inflationary cosmology where the minimally coupled scalar field theory is modified by additional correction terms. Among the most general ten correction terms remarked by Weinberg in context of effective field theory, we consider only two terms, $f_1(\phi) R^2$  and $f_2(\phi) R^{ab}R_{ab}$, following the work by Noh and Hwang where $f_1$ and $f_2$ are constant. The fourth order differential equations for the background universe and the tensor-type perturbation are derived out of this model. We show that these equations can be reduced to second order equations, supposing that $f_n$ are small. From these approximated equations, we find that the propagation speed of gravitational wave is slightly less than the speed of light due to $f_2$ term, and that the evolution of the tensor-type perturbation is conserved in the large scale limit.
\end{abstract}



\maketitle

\section{INTRODUCTION}
For a generic description of the very early universe governed by high energy physics 
where effects of quantum gravity can occur, 
Weinberg \cite{PRD08Weinberg} suggested the most general corrections with four spacetime derivatives, 
$\Delta L$, \bea
&& \Delta L = \sqrt{-g} [ f_1(\phi) (\phi^{,c} \phi_{,c})^2
                          + f_2(\phi) \phi^{,c} \phi_{,c} \Box \phi  
                          + f_3(\phi)(\Box\phi)^2 
						  - f_4(\phi) R^{ab} \phi_{,a} \phi_{,b}
						  	 {}
\nonumber\\
&& \quad\;\, {}	- f_5(\phi) R \phi^{,c} \phi_{,c}	  - f_6(\phi)R\Box\phi 
                          + f_7(\phi) R^2
                          + f_8(\phi)R^{ab}R_{ab} 
 {}
\nonumber\\
&& \quad\;\, {}
                          + f_9(\phi) C^{abcd}C_{abcd}	] 
						  + f_{10}(\phi) \eta^{abcd} C_{ab}^{\;\;\; ef} C_{cdef}   \label{Weinberg Lagrangian} . \eea
Here, $R_{ab}$ is the Ricci tensor, $R$ is the Ricci scalar,  
$\eta^{abcd}$ is a totally antisymmetric Levi-Civita tensor density,
and $C_{abcd}$ is the Weyl tensor.
This $\Delta L$ is added to  
the standard Lagrangian of the minimally coupled scalar field (MSF), $L_{0}$, 
describing the universe filled with scalar field \cite{85Mukhanov, 05Mukhanov, book08Weinberg}, 
given by
\bea
 && L_{0} =  \sqrt{-g} \Big[ {1 \over 16\pi G} R -{1 \over 2} \phi^{,c} \phi_{,c} -V(\phi)  \Big] \label{L0} ,
\eea
where
$G$ is Newton's constant and 
 $V(\phi)$ is a potential as a function of single scalar field $\phi$. 

 These correction terms with just four spacetime derivatives have been previously discussed by Elizalde {\em et al}. \cite{95Elizalde et al} in a different context.
$R^2$ or $R^{ab}R_{ab}$ terms were studied 
by DeWitt (1967) searching for quantum theory of gravity 
and by Birrell and Davis studying on quantum fields in curved space \cite{DeWitt B and D}.
In earlier times, Weyl, Pauli, and Eddington suggested a simpler version of the additional term(s) \cite{WPE}. Especially, the term proportional to $R^2 $, in a pure gravity theory without scalar field, has been discussed by Starobinsky \cite{80 Starobinsky}, a special example of general $f(R)$ gravity \cite{10 Sotiriou and Faraoni, Nojiri:2017ncd, Nojiri:2010wj} which substitutes the standard Einstein-Hilbert action. An inflation model based on Starobinsky gravity as well as non-minimally coupled scalar field theory \cite{13 Kallosh and Linde,11Linde et al, 09Bezrukov and Shaposhnikov, 98 H and N nonMSF} well explains the observational results pictured in the $n_s$(spectral index)-r(tensor-to-scalar ratio) plane, and these are preferred among other inflationary models by Planck Collaboration \cite{Planck2015} who measures the cosmic microwave background (CMB) anisotropy. In addition to inflation, dark energy related scenarios are well accommodated by theories of modified gravity and scalar field \cite{01 H and N, 15Vagnozzi et al, A and T, 06Copeland et al}. Weinberg in his 2008 paper derived the tensor mode equation for only $f_{10}$ correction \cite{PRD08Weinberg}.
 Noh and Hwang considered $f_7$  and $f_8$ as constants without other correction terms and aimed at the explanation of cosmological gravitational wave \cite{97 N and H}. 
Here, we mainly generalize this theory such that $f_7$  and $f_8$ are small corrections as functions of a scalar field.

In section II, we derive gravitational field equations and scalar field equation of motion. In section III, we apply the standard cosmological metric to the equations derived in section II. In section IV, we use a perturbative approximation and obtain solutions under the condition of large scale limit; these are our main results. 
In section V, we briefly discuss our results.
We take the convention of Hawking and Ellis \cite{Hawking and Ellis} and the notation of Hwang and Noh \cite{05 H and N}. Here, $c \equiv 1 \equiv \hbar$ .

\section{EINSTEIN EQUATIONS AND EQUATION OF MOTION WITH TWO CORRECTION TERMS }
The action considered here is 
\bea
 && S    =  \int d^4x \sqrt{-g} \Big[ {1 \over 16\pi G} R - {1 \over 2} \phi^{,c} \phi_{,c} - V(\phi) 
                    + f_1(\phi) R^2   + f_2(\phi)R^{ab}R_{ab}  
           \Big]  \label{action}	,
 \eea
where $f_1(\phi)$ and $f_2(\phi)$ are the dimensionless functions corresponding to $f_7$ and $f_8$ respectively in Eq. (\ref{Weinberg Lagrangian}). 
Varying the action (\ref{action}) with respect to the metric and the scalar field \cite{09Yun, 72Weinberg, book08Weinberg, 83Barth} 
yields the gravitational field equations (GFE) and equation of motion (EOM):
\bea
&&  R_{ab} - {1 \over 2} g_{ab} R - 8\pi G ( T^{(f_1)}_{ab}   + T^{(f_2)}_{ab} )  = 8\pi G T^{(MSF)}_{ab}     
          \label{GFE}   	,
 \eea 
 where \bea	
 && T^{(MSF)}_{ab} = \phi_{,a} \phi_{,b} - \Big( {1 \over 2} \phi^{,c} \phi_{,c} + V \Big) g_{ab}  , 
		  \label{EM tensor MSF}
\\ &&  T^{(f_1)}_{ab} \equiv 2 f_1 \Big( {1 \over 2} R^{2} g_{ab} - 2R R_{ab} - 2 g_{ab} \Box R + 2 R_{;ab}   \Big)  
                                                                   \nonumber \\ && {}   - 8 f_{1,c} R^{;c} g_{ab} + 8 f_{1,(a} R_{,b)} 
                                                                     + 4 f_{1;ab} R - 4 \Box f_1 R g_{ab}  ,
\\ && T^{(f_2)}_{ab} \equiv f_2 g_{ab} R^{cd} R_{cd} - 2g_{ab} (f_2 R^{cd}) _{;cd}  
		      + 4 ( f_2 {R_{(a}}^c ) _{;b)c} - 2 \Box (f_2 R_{ab}) - 4 f_2 {R_a}^c R_{bc}
\nonumber \\ && {} = 2 f_2 \Big( {1 \over 2}  R^{cd} R_{cd} g_{ab} + R_{;ab} - 2 R^{cd} R_{acbd} - {1 \over 2} g_{ab} \Box R - \Box R_{ab}  \Big)  {}
                                              \nonumber \\ &&  {}   + 2 ( -g_{ab} f_{2,c} R^{;c} - 2 f_{2,c} R_{ab;d} g^{cd} + 2 f_{2,c} R^{c}_{(a;b)} + f_{2,(a} R_{,b)}  )  {}
                                                           \nonumber \\ && {}  + 2 ( - f_{2;cd} R^{cd} g_{ab} - \Box f_2 R_{ab} + 2 f_{2;c(a} R_{b)}^{c}    )  \label{EM tensor f2} ,
\eea
and 
\bea && \Box \phi  = V_{,\phi } - f_{1,\phi} R^2  - f_{2,\phi} R^{ab} R_{ab}       \label{EOM}  .
\eea
Here, 
semicolons denote covariant derivatives, symmetrization of a tensor is defined as $T_{(ab)} \equiv {1 \over 2} (T_{ab} + T_{ba})$ , 
d'Alembertian of $\phi$ is written as $ \Box \phi \equiv g^{ab} \phi_{,a;b} $ , 
$V_{,\phi} \equiv {\partial V \over \partial \phi} $, and $ \dot \phi \equiv {\partial \phi \over \partial t} $ .
 In Eq. (\ref{EM tensor f2}), the Bianchi identities \cite{72Weinberg} have been used in order to specify each component of the energy-momentum tensor conveniently. 
 If $f_1$ and $f_2$ are constants, then the GFE are in agreement with the previous results by Noh and Hwang \cite{97 N and H} .

\section{EVOLUTION OF BACKGROUND UNIVERSE AND GRAVITATIONAL WAVE}
 We assume a homogenous, isotropic, and spatially flat Friedmann-Lema\^itre-Robertson-Walker (FLRW) metric \cite{72Weinberg} 
for the description of the background universe and consider tensor-type linear perturbation: 
\bea 
 && ds^2 = a^2 \big[- d \eta^2 + ( \delta_{\alpha \beta}  + 2 C_{\alpha\beta} ) dx^{\alpha} dx^{\beta}    \big]  \label{metric ten PT} .
\eea
Here, $a(t)$ is the cosmic scale factor, $x^0 \equiv\eta $, and $ dt \equiv a d\eta $ .
According to the notation of Hwang and Noh \cite{05 H and N} 
who have formulated cosmological linear perturbation theory in various generalized gravity including scalar- and tensor-type perturbation, 
$ C^{(t)}_{\alpha\beta}$ should be used instead of $C_{\alpha\beta}$ to indicate the tensor mode. 
However, the superscript (t) is omitted in this paper, since we deal with only gravitational wave. 
$ C_{\alpha\beta}({\mathbf x} ,t)$ is tracefree and transverse with respect to
 the flat three-dimensional metric $\delta_{\alpha\beta}$ ,
 $C^{\alpha}_{\alpha} \equiv 0 \equiv C^{\alpha}_{\beta,\alpha}$ .
 $C_{\alpha\beta}$ is also invariant under a gauge transformation 
\cite{80Bardeen, 92Mukhanov, 05Mukhanov, 05 H and N, 11H, 84Kodama}. 
Useful quantities calculated from the metric (\ref{metric ten PT}), are listed in the appendices of Noh and Hwang \cite{97 N and H}. 
They include $G^{a}_{b}, \Box R$, etc.
By substituting the metric  (\ref{metric ten PT}) into GFE (\ref{GFE}) and EOM (\ref{EOM}),
we obtain
\bea 
&& 8\pi G T^{0(MSF)}_{0}
\nonumber \\ &&  = -3 H^2 
- 96 \pi G \big[ (3f_1 +  f_2) ( 2 H \ddot H - \dot{H}^2 + 6 H^2 \dot H  ) 
  + \dot{f}_1 H R + \dot{f}_2 ( 3 H^3  + 2 H \dot H )  \big]    \label{Einstein eq 00-component} , 
\\ && T^{0(MSF)}_{\alpha} = T^{\alpha(MSF)}_{0} = 0 ,
\\ && 8\pi G T^{\alpha (MSF)}_{\beta}
\nonumber \\ && = - ( 2 \dot H + 3 H^2 ) \delta^{\alpha}_{\beta} + D^{\alpha}_{\beta}
\nonumber \\ && - 8 \pi G \Big\{ 4 ( 3 f_1 + f_2 ) \delta^{\alpha}_{\beta} ( 2 \dddot H +12 H \ddot{H} + 9 \dot{H}^2 + 18 H^2 \dot{H} ) 
                                                           + 8 \dot{f}_1 \delta^{\alpha}_{\beta} ( \dot{R} + HR )         
                                                           + 4 \ddot{f}_1 R \delta^{\alpha}_{\beta} 
        \nonumber \\ && {}    - 4 f_1 ( R D^{\alpha}_{\beta} + \dot{R} \dot{C}^{\alpha}_{\beta}  )
                                                            - 4 \dot{f}_1 R \dot{C}^{\alpha}_{\beta}      
       +  2 \dot{f}_2 \delta^{\alpha}_{\beta} ( 8 \ddot{H} + 36 H \dot{H} + 12 H^3   )
                                      +  4 \ddot{f}_2 \delta^{\alpha}_{\beta} ( 2 \dot{H} + 3 H^2 ) 
  \nonumber \\ && {}      +  2 f_2  [ \ddot{D^{\alpha}_{\beta}} + 3 H \dot{D^{\alpha}_{\beta}}
         - 6 ( \dot{H} + H^2 ) D^{\alpha}_{\beta} - { \Delta \over a^2 } D^{\alpha}_{\beta}   
          - 6 ( \ddot{H} + 2 H \dot{H}  ) \dot{C}^{\alpha}_{ \beta } - 4 \dot{H} { \Delta \over a^2 } C^{\alpha}_{ \beta }       ]
 \nonumber \\  && {}    + 2 \dot{f_2} [ 2 \dot{D^{\alpha}_{\beta} } + 3H D^{\alpha}_{\beta}  - 6 ( \dot{H} + H^2 ) \dot{C}^{\alpha}_{ \beta }   ]
    + 2 \ddot{f_2}  D^{\alpha}_{\beta}  \Big\}  \label{Einstein eq spatial component} , \eea 
and
\bea
 &&  \ddot\phi + 3 H \dot\phi + V_{,\phi}
     - 36 f_{1,\phi} \big( \dot{H}^2  + 4  \dot H  H^2 + 4 H^4    \big) 
           -12 f_{2,\phi} \big( \dot H^2 + 3 \dot H H^2 + 3 H^4  \big)                   
= 0     \label{EOM BG}   ,
\eea
where 
the Hubble parameter, $H \equiv {\dot{a} / a}$, the Ricci scalar, $R = 6 ( \dot H + 2 H^2 )$,
 and \bea 
&& D^{\alpha}_{\beta} \equiv  \ddot{C}^{\alpha}_{\beta} + 3H \dot{C}^{\alpha}_{\beta} - { \Delta \over a^2 } C^{\alpha}_{\beta} \label{D def}  .
 \eea 
Putting $f_1$ and $f_2$ to be constant and removing the $\phi$-dependent terms, we get the results which agree with those of Noh and Hwang \cite{97 N and H}.
Therefore, their remarks on the qualitative sameness of the background contribution from  $R^2$ and $R^{ab}R_{ab}$ theories
also hold in this case. 

We can split the energy momentum tensor into the background part (function of only time) 
and the small perturbed part (function of both time and space) 
in the cosmological linear perturbation theory based on the typical FLRW model \cite{11H, 05Mukhanov, book08Weinberg}, 
$T^{a}_{b}({\mathbf x} ,t) =  \overline{T^{a}_{b}}(t) + \delta T^{a}_{b}({\mathbf x} ,t) $.
The background parts are easily read off from the Eqs. (\ref{Einstein eq 00-component}) and (\ref{Einstein eq spatial component}):
\bea
&&  H^2 
+ 32 \pi G \Big[  (3 f_1 + f_2 ) ( 2 H \ddot{H} - \dot H^2  + 6 H^2 \dot H ) 
  + 6 \dot f_1 ( 2 H^3 + H \dot H ) +   \dot f_2 ( 3 H^3 + 2 H \dot H  )    \Big]
  \nonumber \\ && {} = - {  8 \pi G \over 3 } T^{0 (MSF)}_{0}
  = {  8 \pi G \over 3 } \mu^{(MSF)}  = {  8 \pi G \over 3 } \Big( { \dot\phi^2 \over 2 } + V \Big)  ,
\nonumber  \\ && \dot H 
 + 16 \pi G \Big[ 2 (3 f_1 + f_2 ) ( \dddot H  + 3 H \ddot H + 6 \dot H^2    )   
    \nonumber \\ && {}  + 6 \dot f_1 ( 2 \ddot H + 7 H \dot H - 2 H^3  )
                                       + \dot f_2 (  4 \ddot H + 12 H \dot H  - 3 H^3 )
  + 6 \ddot f_1 ( \dot H + 2 H^2 )  +  \ddot f_2 ( 2 \dot H + 3 H^2 )   \Big]  
\nonumber \\ && {}  =  4 \pi G \Big( T^{0(MSF)}_0 - { 1 \over 3 } \overline{T^{\alpha}_{\alpha}}^{(MSF)}   \Big)
     =  -4 \pi G \dot\phi^2 
\label{Friedmann eq}  .
\eea 
The second equation can also be checked by diffentiating the first one and by using the EOM (\ref{EOM BG}).
The perturbed part of Eq. (\ref{Einstein eq spatial component}) is 
\bea
&& D^{\alpha}_{\beta} + 8\pi G \Big\{ 4 f_1 ( R D^{\alpha}_{\beta} + \dot R \dot C^{\alpha}_{\beta}    ) +  4 \dot{f}_1 R \dot{C}^{\alpha}_{\beta}     
     \nonumber \\ && {}  - 2 f_2  [ \ddot{D^{\alpha}_{\beta}} + 3 H \dot{D^{\alpha}_{\beta}}
         - 6 ( \dot{H} + H^2 ) D^{\alpha}_{\beta} - { \Delta \over a^2 } D^{\alpha}_{\beta}   
          - 6 ( \ddot{H} + 2 H \dot{H}  ) \dot{C}^{\alpha }_{ \beta } - 4 \dot{H} { \Delta \over a^2 } C^{\alpha}_{ \beta }       ]
 \nonumber \\  && {}    - 2 \dot{f_2} [ 2 \dot{D^{\alpha}_{\beta} } + 3H D^{\alpha}_{\beta}  - 6 ( \dot{H} + H^2 ) \dot{C}^{\alpha }_{ \beta }   ]
    - 2 \ddot{f_2}  D^{\alpha}_{\beta}       \Big\}  
= 0   \label{GW eq}  .
\eea 
Eq. (\ref{GW eq}) is a fourth order differential equation for $C^{\alpha}_{\beta}({\mathbf x} ,t)$ .
Thus, it is theoretically hard to deal with because more initial conditions are required for numerical analysis and these equations allow unnecessary unphysical solutions. 
With this concern for the problems of higher-derivative theories, 
the research on a perturbative method for reducing the order of derivatives has been done by Simon {\em et al}. \cite{89Simon, 90Simon, 93Parker, 18Solomon}.

\section{second order differentional equations after feedback}
Considering the quantum corrections are small and neglecting $f_n^2$ terms allow the order reduction of the differential Eqs. (\ref{Friedmann eq}, \ref{GW eq}): 
\bea 
&& H^2 = 8 \pi G \Big\{   {1 \over 3  } \mu^{(MSF)} 
 + 8 \pi G ( 3 f_1 + f_2  ) \big[ 8\pi G \big( \mu^{(MSF)} + p^{(MSF)} )^2 + 4 H \dot{p}^{(MSF)}   \big]  
 \nonumber \\ && {}       + 32\pi G H \big[ \dot f_1 \big( 3 p^{(MSF)} - \mu^{(MSF)}  \big) + \dot f_2 p^{(MSF)}    \big]  \Big\}
\nonumber \\ && {} = 8 \pi G \Big\{  { 1 \over 3 } \Big( { \dot\phi^2 \over 2 } + V  \Big)
 - 64\pi G ( 3 f_1 + f_2  ) \Big[  4\pi G \dot\phi^2 \Big( { \dot\phi^2 \over 4 } + V  \Big) +  H \dot\phi V_{,\phi}    \Big]
  \nonumber \\ && {} + 32\pi G H \Big[ \dot{f}_1 \big( \dot\phi^2 - 4V \big) + \dot{f}_2 \Big(  { \dot\phi^2 \over 2 } - V   \Big)  \Big]       
 \Big\}    
\label{Friedmann eq feedback}
\eea
and \bea
 &&  D^{\alpha}_{\beta} + 32\pi G \Big\{  f_1  \dot R \dot C^{\alpha}_{\beta}     +   \dot{f}_1 R \dot{C}^{\alpha}_{\beta}     
  \nonumber \\  && {}       +  f_2  [          3 ( \ddot{H} + 2 H \dot{H}  ) \dot{C}^{\alpha }_{ \beta } + 2 \dot{H} { \Delta \over a^2 } C^{\alpha}_{ \beta }       ]
    + 3 \dot{f_2}   ( \dot{H} + H^2 ) \dot{C}^{\alpha }_{ \beta }   
          \Big\}     = 0  .  \label{GW eq feedback}  
\eea 
A much simplified second order differential equation (\ref{GW eq feedback}) for $C^{\alpha}_{\beta}$ is obtained 
by a feedback method: inserting $D^{\alpha}_{\beta} = \mathcal{O} (f^1_n) $ from Eq. (\ref{GW eq}) 
into the big curly brackets in Eq. (\ref{GW eq}) itself and neglecting very small $ \mathcal{O} (f^2_n)$ terms.
Likewise, using Eq. (\ref{Friedmann eq}) and Eq. (\ref{EOM BG}), we derived a modified Friedmann Eq. (\ref{Friedmann eq feedback}) 
in which the curly brackets may be regarded as ${1 \over 3}$ of the effective energy density in this model.  

Meanwhile, it is allowed to add a term of $f_n^2$-order, $ 96\pi G f_2 (\dot H + H^2) D^{\alpha}_{\beta} $ , to Eq. (\ref{GW eq feedback}) and to recover the $f_1$ gravity terms before the feedback:
\bea
&& D^{\alpha}_{\beta} + 32 \pi G \Big\{   \big( f_1  R \big)  \dot{\phantom{i}} \dot C^{\alpha}_{\beta} +      f_1 R D^{\alpha}_{\beta}
  + 3 \big[ f_2  ( \dot H + H^2 ) \big] \dot{\phantom{i}}  \dot{C}^{\alpha }_{ \beta } 
                                            + 3 f_2  ( \dot H + H^2 )  D^{\alpha}_{\beta}  + 2  f_2 \dot{H} { \Delta \over a^2 } C^{\alpha}_{ \beta }     \Big\}
\nonumber \\  && {}  = F D^{\alpha}_{\beta} + \dot F \dot{C}^{\alpha}_{\beta} + 64\pi G f_2 \dot H { \Delta \over a^2 } C^{\alpha}_{ \beta }
= 0 \label{GW eq feedback 2} , 
\eea
where \bea
F \equiv  1 + 32\pi G [ f_1 R + 3 f_2 ( \dot H + H^2 ) ]  .  \eea
Dividing Eq. (\ref{GW eq feedback 2}) by $F$ and using the definition of $D^{\alpha}_{\beta}$ in Eq. (\ref{D def}) 
lead to an equation for the tensor mode in the compact form: 
\bea
&& {1 \over a^3 F } \big(  a^3 F  \dot C^{\alpha}_{\beta}   \big)\dot{\phantom{i}}
 - \big( 1- 64\pi G f_2 \dot H   \big) { \Delta \over a^2 } C^{\alpha}_{ \beta }
   \nonumber \\ && {} =  { 1 \over {a^2 z} } \Big[ { v^{\alpha}_{\beta}}'' 
 - \Big( { z'' \over z } +  c^2_T \Delta   \Big) v^{\alpha}_{\beta}    \Big]   = 0   \label{MS eq} ,
\\ && {}  v^{\alpha}_{\beta} \equiv z C^{\alpha}_{\beta} ,
          \quad      z \equiv a \sqrt{F}  , \eea 
and \bea
 && {} c^2_T \equiv 1- 64\pi G f_2 \dot H . 
\eea 
Here, $ ' \equiv {\partial \over \partial\eta} $.
Eq. (\ref{MS eq}) is often called Mukhanov-Sasaki equation \cite{85Mukhanov, 92Mukhanov, 86Sasaki}.
If $c_T  $ is the gravitational wave propagation speed,
then it is affected not by the general function $f_1(\phi)$, but by the small $f_2$ correction term depending on time.
Moreover, $c_T$ should be less than the speed of light, thus the constraint that $ f_2 \dot H > 0 $  is required.

In the large scale limit, a general integral form solution is obtained:
\bea
 C^{\alpha}_{\beta}({\mathbf x} ,t)     =  c^{\alpha}_{\beta}({\mathbf x})  + d^{\alpha}_{\beta}({\mathbf x})   \int^{t} {dt \over {a^3 F}} 
\label{LS sol}  , 
\eea
where $ c^{\alpha}_{\beta}(\mathbf{x})  $  
and $ d^{\alpha}_{\beta}(\mathbf{x}) $ 
 are the time-independent integration constants.
Ignoring the decaying transient  $d$-solution in an expanding universe,
we note that
the evolution of tensor type perturbation in the large scale limit 
is described by the conserved quantity $ c^{\alpha}_{\beta}(\mathbf{x})  $.

\section{DISCUSSIONS}
 We have derived complicated fourth order differential equations of the gravitational wave as well as the background evolution in the inflationary universe implemented with the additional two modified gravity theories including a scalar field. Reducing the order by the perturbative approximation yields the more tractable equation and its solutions in the large scale limit. With model-dependent variables $F, z$, or $c_T$ \cite{13 Yunes and Siemens, 10Garfinkle et al, 17 Creminelli and Vernizzi} , the form of Eq. (\ref{MS eq}) is maintained in various generalized gravity theories such as  a model motivated by string theory.
Those variables have been tabulated in Ref. \cite{05 H and N}    . 
If $f_1$ and $f_2$ are constants,
Einstein gravity and Starobinsky gravity correspond to a limit of $F =1$ and $ F = 1 + 32\pi G f_1 R$  respectively . 
 It would be more appropriate to call Eq. (\ref{MS eq}) Field-Shepley \cite{68Field} equation if the priority were concerned.

According to Weinberg \cite{PRD08Weinberg} , if the field equations derived from the MSF Lagrangian (\ref{L0}) are used in the correction Lagrangian (\ref{Weinberg Lagrangian}) and $\phi$ and $V(\phi)$ are suitably redefined, then Eq. (\ref{Weinberg Lagrangian}) can be simplified to have only three terms, $f_1, f_9$, and $f_{10}$ . In other words, the ten terms in Eq. (\ref{Weinberg Lagrangian}) are not independent to one another if the perturbative method at the action level and the redefinition approach are applied.
We suggest an interpretation of the logic behind his argument that is simpler than our approach to the full Lagrangian as follows. Assuming that $\Delta L$ (\ref{Weinberg Lagrangian}) is much smaller than $L_0$ (\ref{L0}), Einstein's equation (we set $8 \pi G \equiv 1$ in this section only)
\bea
R_{ab} = \phi_{,a} \phi_{,b} + g_{ab} V
\eea 
derived from $L_0$ (\ref{L0})
and its trace equation
\bea
R = 2 ( X + 2 V )
\eea 
with a convenient definition $X \equiv {1 \over 2}  g^{ab} \phi_{,a} \phi_{,b} $ 
can be put into $\Delta L$ (\ref{Weinberg Lagrangian}).
Assuming that $ f_8 = -4 f_7 $,
\bea
  f_7 R^2 + f_8 R^{ab} R_{ab} = - 12 f_7 X^2 \equiv 4 f_1 X^2 .
\eea  
Thus, the $f_1$-gravity form
\cite{99  AP D and M}
is obtained from the seventh and eighth terms in $\Delta L$ (\ref{Weinberg Lagrangian}) with the abovementioned assumptions.  
 Our approach in a different context results in a modified propagation speed of gravitational wave that is measurable in principle. 
 We selected and considered only two terms, $f_7$ and $f_8$ in the correction Lagrangian (\ref{Weinberg Lagrangian}) 
and directly analyzed the action without any redefinitions and simplification, while we and Weinberg share the same assumption that the correction Lagrangian is small.
We used the approximation at the wave equation (\ref{GW eq}), while he did the approximation at the action level. 
Comparison between two methods may be another issue.  

There are several future investigations about this research.
Firstly, quantizing Eq. (\ref{MS eq}) from the action level is straightforward 
by following the known prescriptions \cite{05 H and N, 92Mukhanov} .
The unitarity shall be considered during quantization of the theories here to preserve the inner product of quantum states; however, the unitarity-violating term is encountered in a study of quantum cosmology \cite{12 Kiefer and Kraemer} . Indeed, quantizing gravity is an abstruse issue for the very early universe. 
More fundamentally, various generalized gravity theories with higher-derivative expansion are motivated by string theory 
\cite{12EMM, 07Gasperini, 03 Gasperini and Veneziano, 87GSW} .  
Secondly, if the Riemann-tensor-squared Lagrangian is studied, 
then the tensor mode equations in this paper will be able to transform into Weinberg's counterpart \cite{PRD08Weinberg}.
Thirdly, a heavy numerical analysis may allow a comparison of the exact equations and the approximate equations.  

\begin{acknowledgments}
The author is grateful 
to Prof. Jai-chan Hwang for his teachings on cosmology
and 
to Prof. Sang Gyu Jo for his thoughtful advice and critical review.
The author also thanks Prof. Chan-Gyung Park for his much help in {\em Mathematica} usage.
\end{acknowledgments}

\end{document}